\documentclass[a4paper,fleqn,usenatbib]{mnras}

\usepackage{graphics,epsf}
\usepackage{amsmath}                
\usepackage{amsfonts}               
\usepackage{amssymb}                
\usepackage{epsfig}                 
\usepackage{graphicx}

\def \msyr{~\rm{M_{\odot}}~\rm{yr^{-1}}}
\def \cm{~\rm{cm}}
\def \s{~\rm{s}}
\def \km{~\rm{km}}

\def \K{~\rm{K}}
\def \g{~\rm{g}}

\def \AU{~\rm{AU}}

\def \yr{~\rm{yr}}

\def \rmModot{~\rm{M_\odot}}

\usepackage{xcolor}
\definecolor{redak}{rgb}{0.9,0.15,0.05}
\def \textred{}

\title[3D simulations of GEE]{Simulating the Onset of Grazing Envelope Evolution of Binary Stars}

\author[S. Shiber et al.]{
Sagiv Shiber,$^{1}$\thanks{E-mail: \href{shiber@campus.technion.ac.il}{shiber@campus.technion.ac.il}}
Amit Kashi,$^{2,1}$\thanks{E-mail: \href{mailto:kashi@ariel.ac.il}{kashi@ariel.ac.il}}
and Noam Soker$^{1}$\thanks{E-mail: \href{mailto:soker@physics.technion.ac.il}{soker@physics.technion.ac.il}}
\\
$^{1}$Physics Department, Technion -- Israel Institute of Technology, Technion City -- Haifa 3200003, Israel\\
$^{2}$Physics Department, Ariel University, Ariel, POB 3, 40700, Israel
}

\date{Accepted XXX. Received YYY; in original form ZZZ}

\pubyear{2016}

\begin{document}
\label{firstpage}
\pagerange{\pageref{firstpage}--\pageref{lastpage}}
\maketitle

\begin{abstract}
We present the first three-dimensional gas-dynamical simulations of the grazing envelope evolution (GEE) \textred{{{{{{of stars}}}}}}, with the goal of exploring the basic flow properties and the role of jets at the onset of the GEE.
In the simulated runs, a secondary main-sequence star grazes the envelope of the primary asymptotic giant branch (AGB) star.
The orbit is circular at the radius of the AGB primary star on its equator.
We inject two opposite jets perpendicular to the equatorial plane from the location of the secondary star, and follow the evolution for several orbital periods.
We explore the flow pattern by which the jets eject the outskirts of the AGB envelope.
After one orbit the jets start to interact with gas ejected in previous orbits and inflate hot low-density bubbles.
\end{abstract}

\begin{keywords}
binaries: close --- stars: AGB and post-AGB --- stars: winds, outflows --- ISM: jets and outflows
\end{keywords}

\section{INTRODUCTION}
\label{sec:intro}

Numerical hydrodynamical simulations of the common envelope evolution (CEE) have been performed for thirty years, e.g., the 2D simulations conducted by \cite{BodenheimerTaam1984}, followed by the 3D simulations conducted by \cite{LivioSoker1988} and then by \cite{RasioLivio1996}.
More sophisticated simulations have followed  (e.g., \citealt{SandquistTaam1998, Lombardi2006, DeMarcoetal2011, Passy2011, Passyetal2012, RickerTaam2012, Ohlmannetal2016, Iaconietal2016, Staffetal2016, IvanovaNandez2016, Kuruwitaetal2016, NandezIvanova2016}).
Two main goals are behind many of these simulations. The first goal is to determine the final orbital separation between the core of the giant star and the more compact companion (secondary) star. The second goal is to understand the manner by which the envelope is removed.
Although in many cases less than the energy released by the in-spiraling binary system is required to unbind the CE (e.g., \citealt{DeMarcoetal2011, NordhausSpiegel2013}), the removal of the CE is not free from problems (e.g., \citealt{DeMarcoetal2011, Passy2011, Passyetal2012, RickerTaam2012, Soker2013, Ohlmannetal2016}).

To overcome the envelope-removal obstacle, extra energy sources have been proposed, such as recombination energy (e.g., \citealt{Nandezetal2015} for a recent paper), and jets launched by the secondary star (e.g., \citealt{Soker2004}). Also, \cite{Soker2004} suggested that in some cases, both for stellar and sub-stellar companions, the energy source might be the giant luminosity itself (nuclear burning in the core), under the  assumption that fast rotating envelopes efficiently form dust and have high mass-loss rates.

We adopt the view that in many cases jets launched by the companion can facilitate CE removal (e.g., \citealt{Soker2014}). This holds for main-sequence (MS) stars as well \citep{SchreierSoker2016}, as they can accrete mass at a high rate \citep{Shiberetal2015}. \cite{ArmitageLivio2000} and \cite{Chevalier2012} studied CE ejection by jets launched from a neutron star companion, but they did not consider jets to be a general CE ejection process. When the jets become efficient in envelope removal, such that they remove the entire envelope outside the orbit of the companion, the system does not enter a CE phase. Instead, the system experiences the grazing envelope evolution (GEE; \citealt{SabachSoker2015, Soker2015, Soker2016a, Soker2016b}).

Our goal is to explore some basic properties of the flow at the start of the GEE. The numerical set up is described in section \ref{sec:numerics}, and the results are presented in section \ref{sec:results}. We summarize in section \ref{sec:summary}.

\section{NUMERICAL SET-UP}
\label{sec:numerics}

We run the stellar evolution code \texttt{MESA} \citep{Paxtonetal2015} to obtain an asymptotic giant branch (AGB) model
with zero-age-main-sequence mass of $M_{\rm ZAMS}=3.5 \rmModot$, solar metallicity, and no rotation.
We let the star evolve until it reaches the AGB stage after $3 \times 10^{8}$ years.
At that point the stellar mass is $M_{\rm AGB} \simeq 3.4 \rmModot$,
The radius is $R_{\rm AGB} \simeq 1 \AU$ and its effective temperature is $T_{\rm eff} \simeq 3\,400 \K$.
We then import the AGB model into the hydrodynamic code \texttt{FLASH} \citep{Fryxell2000}, and
prepare the setup for our GEE simulation.
We employ a uniform Cartesian grid with a cell size of $4.6875 \times 10^{11} \cm$, and position the AGB at the center $(x,y,z)=(0, 0,0)$.
The grid is taken as a cube with side length of $1.2 \times 10^{14} \cm$.
We use an equation of state of an ideal gas with adiabatic index $\gamma = 5/3$.
\textred{{{{{{ This neglects radiative cooling that might be significant in regions close to the surface, where the photon outward diffusion time is short.  }}}}}}

As we are not interested in the inner parts of the AGB star that requires short time steps, its inner $0.33 \AU$ region was replaced with a constant density sphere. This artificial setting causes some numerical mass and energy outflow from the inner region.
To verify numerical stability, we run the simulation for two dynamical times before inserting the secondary.
The AGB stellar model develops a weak outflow, with kinetic energy negligible relative to the kinetic energy of the jet we will insert later.

We insert a low-mass MS secondary star in a Keplerian orbit around the AGB star on its equator, at a radius of {{{{$1.5 \times 10^{13} \cm$}}}} in two cases and at a radius of {{{{$1.3 \times 10^{13} \cm$}}}} at two other cases.
{{{{We keep the radius constant, as under the GEE prescription the spiral-in process is slow, or does not exist even. }}}}
{{{{{ We note that in our simulations we do not consider the deformation of the primary stellar structure as a result of the secondary star. As it spirals-in the secondary star will spin-up the envelope of the primary star. This will enhance the equatorial radius, and will ease the removal of mass from the envelope. }}}}}

We neglect the effect of losing primary mass and envelope deformation, and keep primary gravity constant and spherically symmetric during the entire simulation. The gravity from the secondary MS star is neglected in solving the hydrodynamic equations.
{{{{This is justified for two reasons. First, the mass of the secondary star is much smaller than that of the primary star, so the secondary contribution to the binding energy of the envelope is small. Secondly, the jets deposit their energy to the surrounding gas far from the secondary location. In that region the secondary gravity is smaller than the primary gravity. }}}}

The secondary star is assumed to launch bipolar jets with a half-opening angle of $30^\circ$ and a velocity of $700 \km \s^{-1}$.
{{{{{  This is 1.2 the escape speed from a low mass main sequence star. As it is usually done in such simulations of jets, the terminal velocity takes into account the action of the secondary stellar gravity and the acceleration mechanism of the jets. }}}}}
The jets are continuously injected from the momentarily location of the secondary star, and also have azimuthal velocity component equals to the orbital velocity of the secondary star, which is about $55 \km \s^{-1}$. The jets are numerically inserted in a cone of $10^{12} \cm$ length,
which means a length of two cells in each direction.

The amount of mass injected by the two opposite jets was calculated assuming Bondi-Hoyle-Lyttleton (BHL)
accretion rate (\citealt{Bondi1952})
\begin{equation}
\dot{M}_{{\rm BHL}}=\pi v_{r}\rho_{e} R_{{\rm acc}}^{2} = \pi v_{r}\rho_{e}
 \left( \frac{2GM_s}{v_r^{2}+c_s^{2}} \right)^2 ,
\label{eq:dotmaccrete}
\end{equation}
where $c_s$ is the sound speed and $v_r$ is the relative velocity between the secondary star
and the envelope.
For a companion performing a Keplerian motion on the
cool outskirts of an AGB star $v^2_r \gg c^2_s$, and the accretion radius is
\begin{equation}
R_{\rm acc} \simeq \frac{2GM_{\rm s}}{GM_{\rm AGB}/R_{\rm AGB}}=2\frac{M_{\rm s}}{M_{\rm AGB}}R_{{\rm AGB}}.
\label{eq:racc}
\end{equation}
\textred{{{{{ We took $\rho_e$ to be the density at the secondary star location. For practical reasons, we took it as the density in the last shell of the imported \texttt{MESA} model, $\rho_e \simeq 4 \times 10^{-9} \g \cm^{-3}$. For a secondary mass of $M_{\rm s}=0.5 \rmModot$, this gives an accretion rate of $\dot{M}_{{\rm BHL}} \simeq 0.02 \msyr $. We can take the mass that is accreted to be the mass that enters the accretion cylinder, namely, the mass which its impact parameter from the secondary star is $b<R_{\rm acc}$. This implies that the mass being accreted in one orbit is the mass residing in a torus which its major radius is the orbital separation and its minor radius is the accretion radius. In that case we found the  accretion rate to be $0.1 \msyr$. Namely, equation (\ref{eq:dotmaccrete}) already considers that the mass accretion rate is much below that of BHL, at the surface of the star, by a factor of about five in our case. Moreover, we expect the secondary star to raise a tidal bulge at its location on the surface. This will further increase the mass accretion rate. Over all, we take an accretion rate that is about one order of magnitude below that expected by simple theoretical arguments, but one that might describes the real situation. }}}}}

{{{{ In their simulation of the CEE, \cite{RickerTaam2012} find the mass accretion rate onto the secondary star to be much smaller than the BHL value. However, they do not include jets that can remove angular momentum and high-entropy gas from the regions around the secondary star.
Without a process that removes energy, the high pressure that is built by the accreted gas prevents further accretion and the formation of an accretion disc or an accretion belt (e.g. \citealt{MacLeodRamirezRuiz2015}).
It is anticipated that when jets are launched and remove angular momentum and high-entropy gas, the accretion rate will be closer to the BHL prescription \citep{Shiberetal2015, Staffetal2016MN}.
The accretor can then launch $10-40$ per cent of the accreted mass through the jets, as it is found for jets from YSOs (e.g., \citealt{Pudritzetal2012, Federrathetal2014}). }}}}
{{{{{ In any case, as discussed above, equation (\ref{eq:dotmaccrete}) already takes into account highly inefficient accretion. We further adopt a conservative approach, and take the jets to carry out only about $13 \%$ of the accretion energy. We do this by taking a mass loss rate onto the two jets together  }}}}}
\begin{equation}
\dot{M}_{\rm jet} =  10^{-3}  \msyr \simeq 0.05 \dot{M}_{\rm BHL}  ,
\label{eq:dotmjet}
\end{equation}
{{{{{ As a further cautionary step, we also simulated two cases with five times lower jets power, by taking
 $\dot{M}_{\rm jet} = 0.2\times 10^{-3} \msyr $. }}}}}

\section{RESULTS}
\label{sec:results}

\begin{figure*}
\centering
\includegraphics[trim= 1cm 0.4cm 0.2cm 0.3cm,clip=true,width=0.9\textwidth]{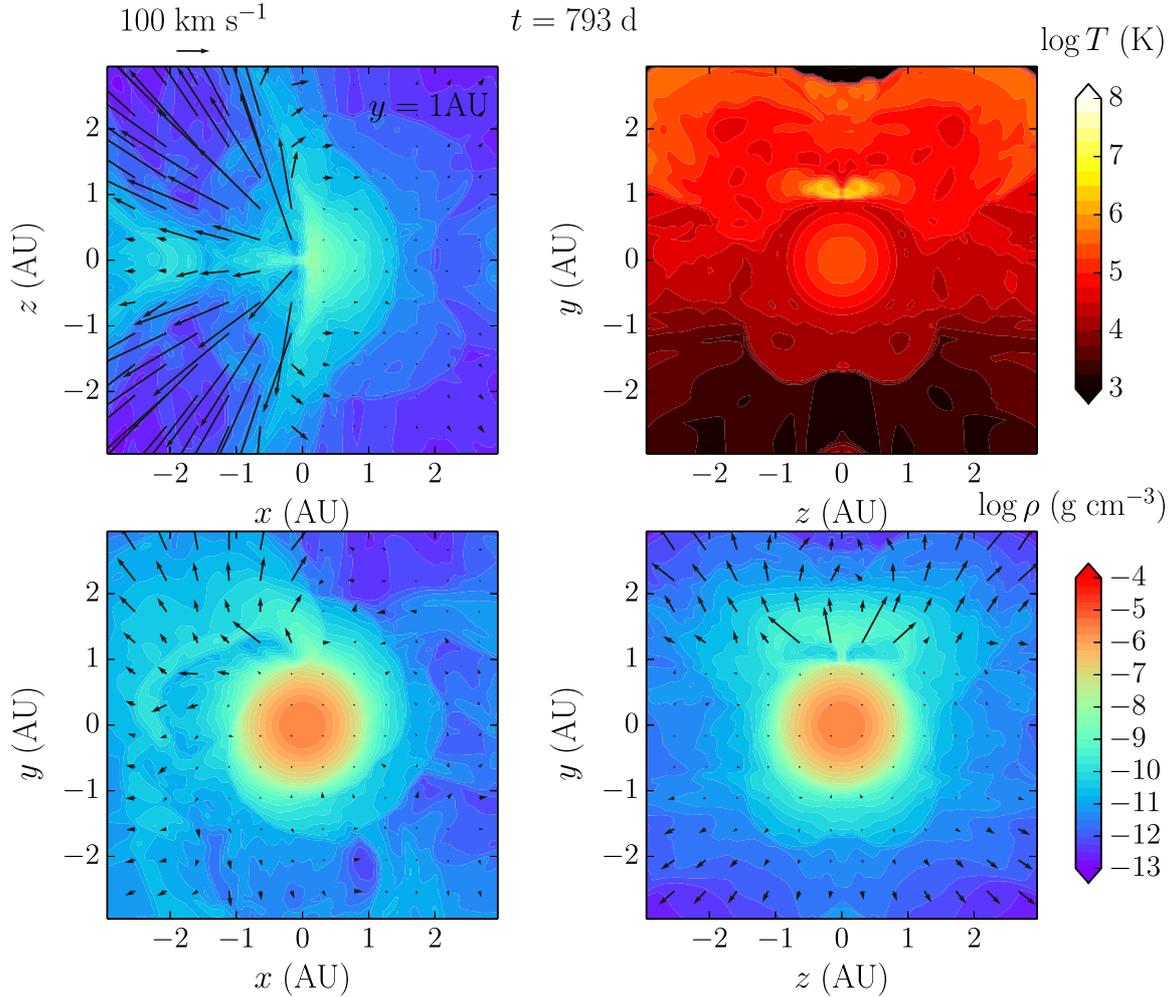} 
\caption{The hydrodynamic properties of Run 1 ($\dot{M}_{\rm{jet}}=10^{-3} \msyr$; $P=198.4 \; \rm{days} $) after four orbital periods.
The companion is at $(x,y,z)=(0,1~ \rm{AU},0)$ on every panel.
The plane $x-y$ is taken to be the equatorial plane.
The upper-left panel shows the density and velocity in the $x-z$ plane at $y=1~ \rm{AU}$, namely, goes through the secondary location in perpendicular to the orbital plane.
The lower-left panel shows density and velocity in the orbital plane.
The lower-right panel shows the density and velocity in the $z-y$ plane, a plane cutting the centers of the two stars.
The upper-right panel shows the temperature in the same plane.
}
\label{fig:4panel}
\end{figure*}

We run four simulations differing in the binary separation and mass injection rate of the jets, as summarized in Table \ref{tab:simulationsummary}.
The jets expel gas from the outer layers of the AGB star, causing mass interior to the location of the secondary to flow out.
Fig. \ref{fig:4panel} shows slices of the logarithm of density, and the absolute value of the projected velocity vector through the orbital and perpendicular planes at the time the companion in Run 1 ends four orbital periods.
Slices of the logarithm of temperature through the perpendicular plane are also shown.
The jets interact with AGB envelope gas in the vicinity of the secondary star, and with circum-stellar gas that was expelled in previous orbits. The flow structure becomes very complicated.
\begin{table}
    \begin{tabular}{l l l l l}
     \hline
     \hline
        Run       & $\dot M_{\rm jet}$					& Companion                & Ejected                & Unbound              \\
        \#        & ($\rm{M_{\odot}}~\rm{yr^{-1}}$)       & distance                 & mass                   & mass                 \\
                  &                                       & (cm)                     & ($\rm{M_\odot}$)       & percentage           \\
\hline
        1            & $10^{-3}$                          &  $1.5 \times 10^{13}$    & 0.048                  & 93.6                 \\
        2            & $2\times 10^{-4}$                  &  $1.5 \times 10^{13}$    &                        &                      \\
        3            & $10^{-3}$                          &  $1.3 \times 10^{13}$    & 0.241                  & 87.8                 \\
        4            & $2\times 10^{-4}$                  &  $1.3 \times 10^{13}$    &                        &                      \\
     \hline
     \hline
    \end{tabular}
\caption{List of simulations.
}
\label{tab:simulationsummary}
\end{table}

\begin{figure}
\includegraphics[trim= 0.0cm 0.5cm 0.0cm 1.0cm,clip=true,width=0.95\columnwidth]{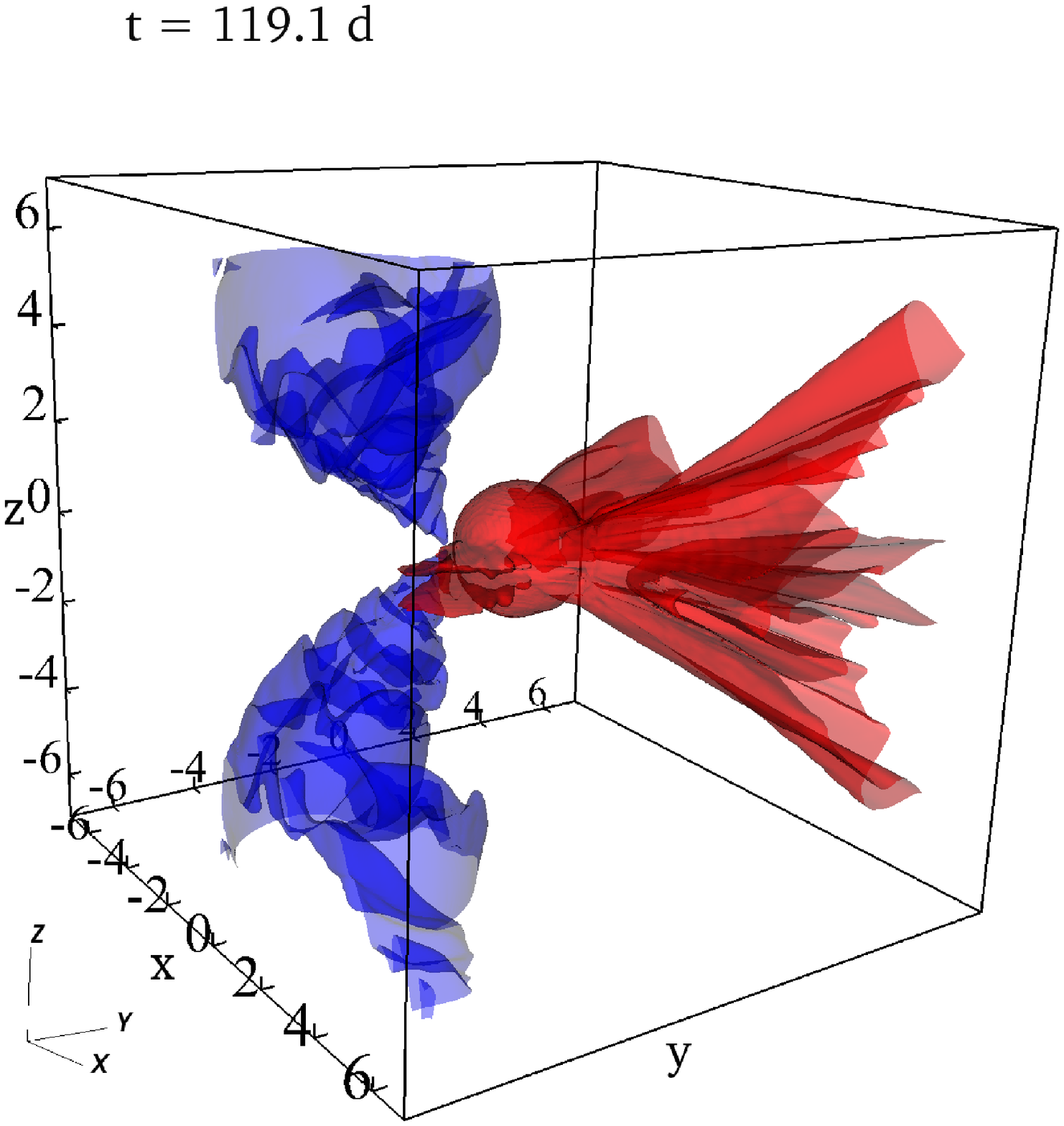}  
\caption{3D map of contour surfaces with $99.5$ per cent of AGB gas (red) and of $50$ per cent of jet gas (blue), after one orbital period for Run 1.
Axes run from $-7\times 10^{13} \cm$ to $7\times 10^{13} \cm$.
The orbital plane is along the line of sight and the secondary star is on the left side of the AGB surface.}
\label{fig:tracers}
\end{figure}

With \textit{MASS\_SCALAR} option in \texttt{FLASH}, we use two `tracers' to follow the material originated from the AGB envelope and the gas injected by the jets.
The tracers indicate the fraction of mass originate from the AGB star and from the jets in each cell.
Fig. \ref{fig:tracers} presents a 3D map of contour surfaces with $99.5$ per cent of AGB gas (red) and of $50$ per cent of jet gas (blue), at $t = 119~{\rm days}$ for Run 1.
The morphology of the jets can clearly be seen in the figure.
The jets are injected from the secondary star in two opposite directions perpendicular to the orbital plane (they also have an azimuthal velocity component due to the orbital motion). The jets are immediately diverted outwards by the dense AGB gas.
%

\begin{figure}
\includegraphics[trim= 0.0cm 0.0cm 1.2cm 0.0cm,clip=true,width=0.95\columnwidth]{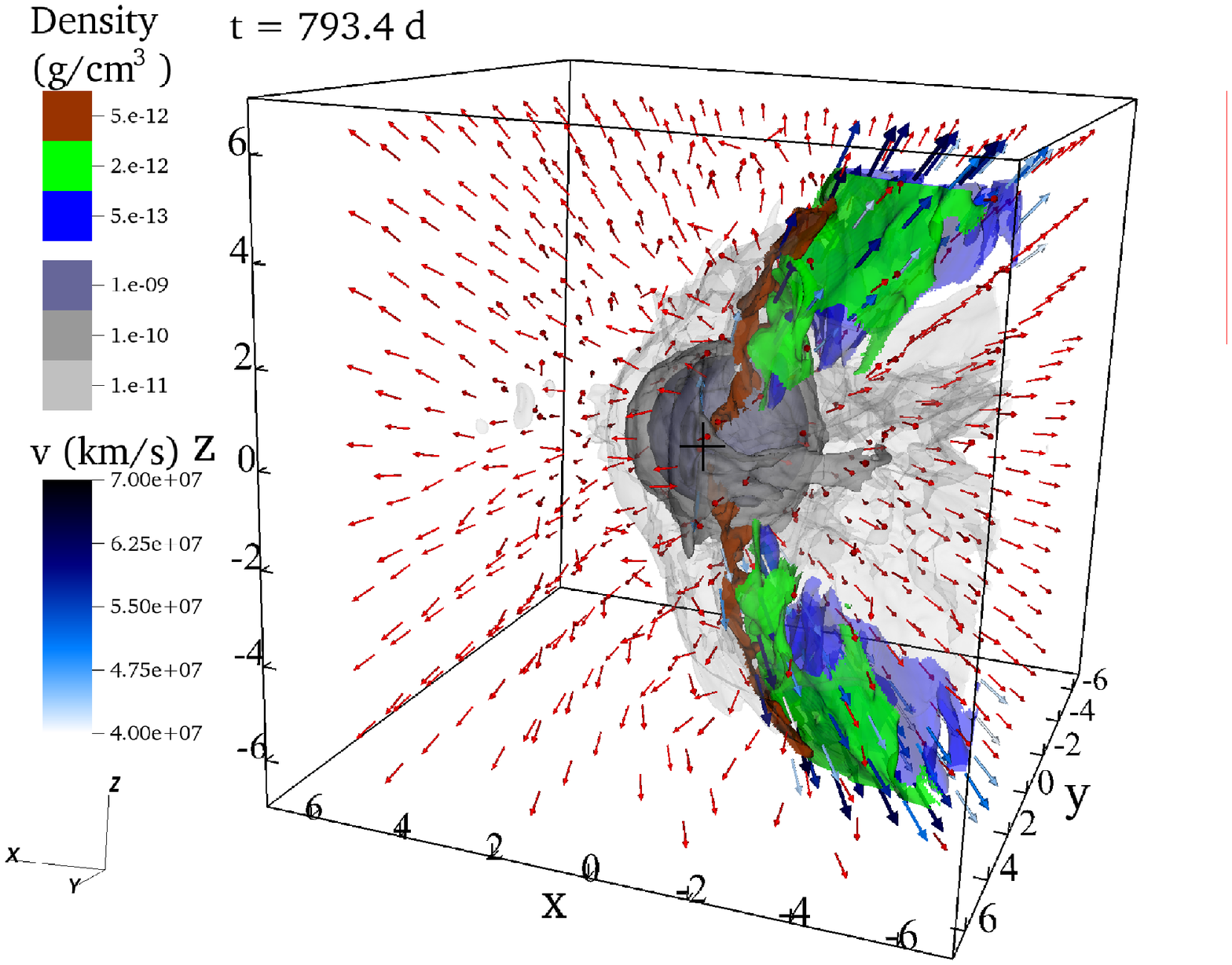} 
\caption{
The flow structure after four orbital periods in Run 1.
Axes run from $-7\times 10^{13} \cm$ to $7\times 10^{13} \cm$.
The AGB is at the center and the secondary is marked by the ``$+$'' sign. The secondary is moving to the left in the figure.
The red-green-blue colors show gas with velocities larger than $400~\rm{km~s^{-1}}$, that originated in the jets and got deflected.
The gray colors show slower gas and at higher densities,
corresponding to gas originated from the AGB envelope.
The blue-scale velocity arrows are
limited to a minimal velocity of $400~\rm{km~s^{-1}}$ and their size and color are proportional
to the module of the velocity.
The black arrows are of uniform size and represent velocities smaller than $400~\rm{km~s^{-1}}$.
}
\label{fig:three-dimensional_density}
\end{figure}

In Fig. \ref{fig:three-dimensional_density} we present a 3D map of density contour surfaces.
Arrows represent the flow velocity.
This figure clearly emphasizes the complicated flow structure.
The figure shows separate color schemes and vector properties for velocity above and below $400~\rm{km~s^{-1}}$.
The high velocity gas comes from the jets and the AGB layers it collided with.
It can be clearly seen that the jets are deflected towards lower density
regions.
The slow velocity gas shows both rotation as a result of the secondary orbit,
and an outflow as a result of the jets.
%

\begin{figure}
\includegraphics[trim= 0.0cm 0.6cm 0.0cm 0.4cm,clip=true,width=0.99\columnwidth]{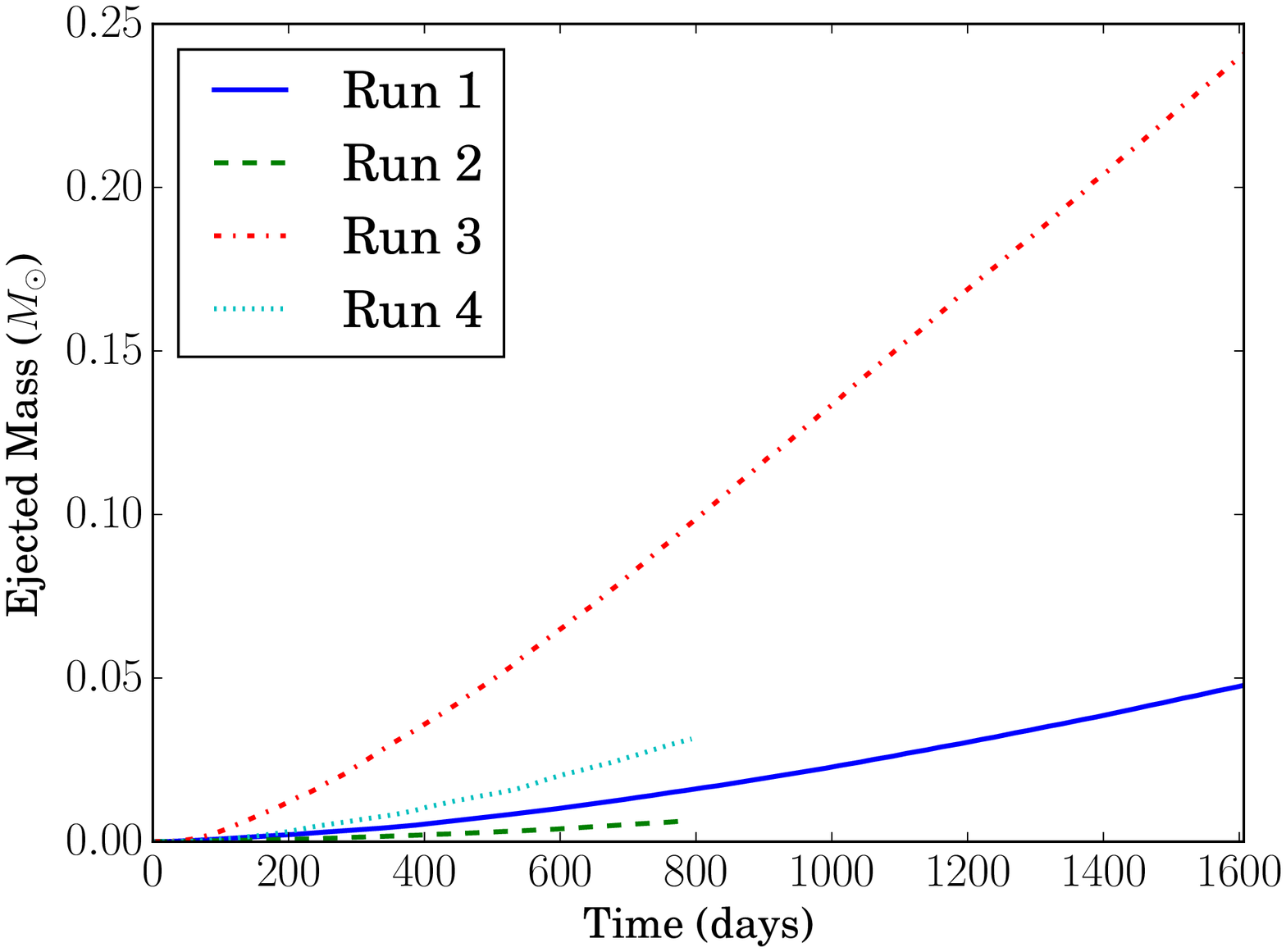}
\caption{The ejected mass as function of time in our four simulations. The ejected mass is the mass that crosses out through a spherical shell of radius $4 \AU$. }
\label{fig:logMout}
\end{figure}

We measured the mass that left the system through a sphere near the edge of the grid.
Fig. \ref{fig:logMout} shows this ejected mass as a function of time for the four simulations.
In Runs 3 and 4 the secondary star starts its circular orbit inside the envelope, and larger amount of mass is ejected from the system during the time of the simulations. Higher values of $\dot{M}_{\rm jet}$ also yielded higher value of ejected mass. Both these results are according to expectations.
In late times of $t \ga 300$~days, the time dependence is close to being exponential with similar rate for all four simulations.
Most of the mass injected in the jets leaves the grid.
The ejected mass is up to $0.1 \rmModot$ at $t=800$~days, which is $\approx10 \textrm{--} 50$ times the mass injected in the jets.
In Run 1 we find that $94$ per cent of the mass that left the grid has a positive energy, namely it is unbound.
In Run 3 we find that $88$ per cent is left unbound.
%

\begin{figure}
\includegraphics[trim= 0.0cm 0.6cm 0.0cm 0.4cm,clip=true,width=0.99\columnwidth]{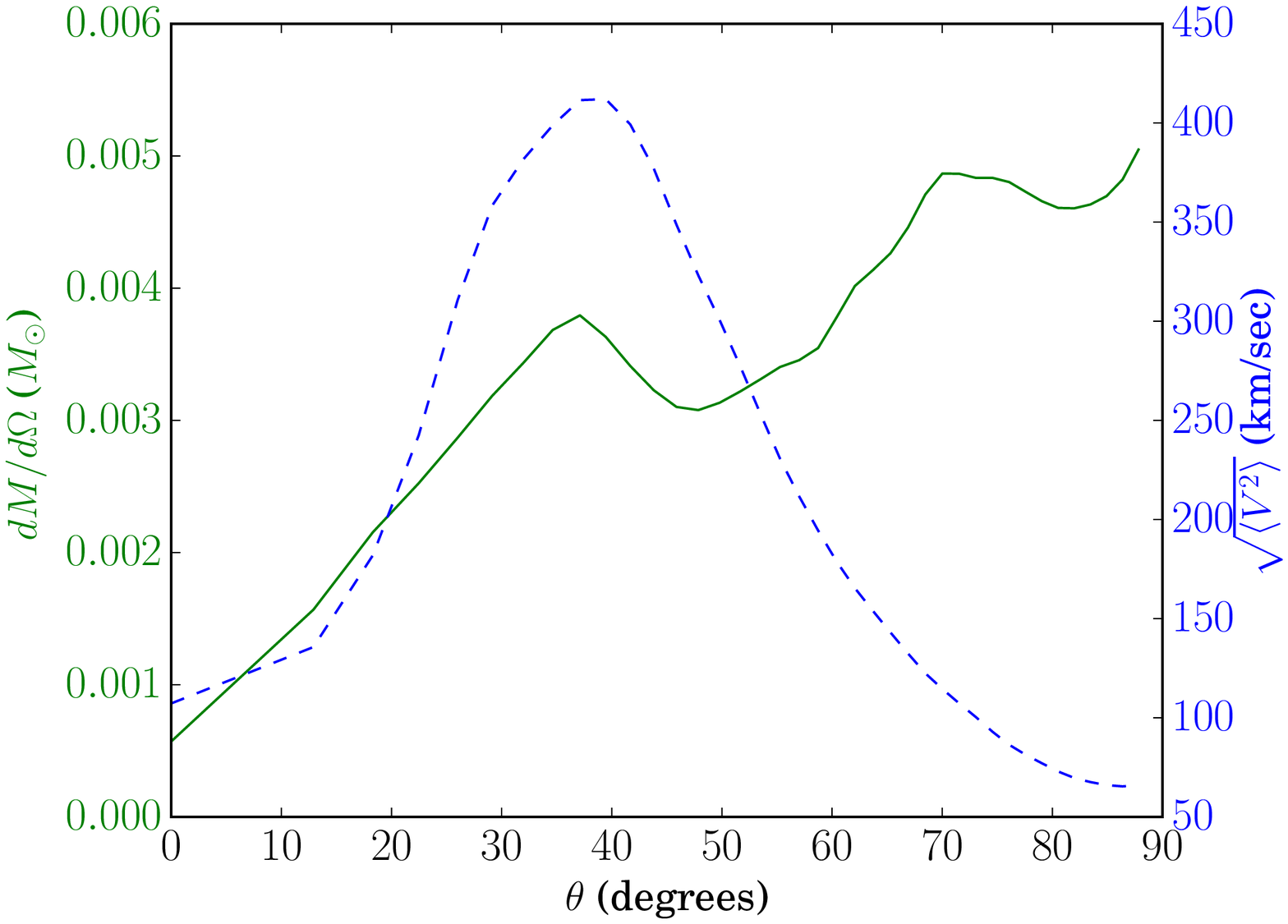}
\caption{The mass lost from the grid per unit solid angle (green line) and $\sqrt{<v^2>}$ (dashed blue line) as a function of the polar angle $\theta$,
for the first $4.4 \yr$ of the simulation of Run 1.}
\label{fig:angle}
\end{figure}

In Fig. \ref{fig:angle} we present the mass that left the grid over the first 8 orbital periods per unit solid angle as a function of the angle from the pole ($\theta=0^\circ$ is the polar direction and $\theta =90^\circ$ is the equatorial plane). The graph represents the two sides of the equatorial plane as they are symmetric.
We also show the value of {{{{$\sqrt{<v^2>} \equiv \sqrt{ 2 dE_k(\theta) / dM(\theta)}$}}}}, where
$d E_k(\theta)$ and  $d M(\theta)$ are the kinetic energy and mass, respectively, that left the grid through a circular surface from $\theta$ to $\theta + d \theta$.
The quantities were calculated from $t=0$ to $t=4.4 \yr$, about 8 orbital periods.
%

From Fig. \ref{fig:angle} we learn the following. (1) There is a concentration of outflow in the equatorial plane (see also Fig. \ref{fig:three-dimensional_density}).
This equatorial outflow is achieved without any gravitational field of the secondary star and without any rotation of the primary AGB star.
The equatorial outflow is a result of the action of the jets.
(2) Along  $\simeq 35^\circ$--$45^\circ$ from the polar directions there is a fast relatively massive outflow.
This is mainly due to the material that originated in the jets and is deflected by the AGB envelope (see also Figs. \ref{fig:tracers} and \ref{fig:three-dimensional_density}).
The outflow geometry and its role in shaping the descendent nebula is a subject of a forthcoming paper.

\section{SUMMARY}
\label{sec:summary}

We conducted preliminary three-dimensional hydrodynamical simulations of the onset of the GEE.
The GEE is based on jets launched from the secondary MS star that efficiently remove mass from the envelope of the primary AGB star,
to the extent it prevents the formation of a common envelope.
In the GEE the secondary star grazes the surface of the giant star, as in Run 1 and Run 2 listed in Table \ref{tab:simulationsummary}.
For comparison we also simulated two cases where the secondary star starts rotating inside the giant envelope (Run 3 and Run 4).

As expected, the flow structure is complicated, e.g., the outflow strongly depends on latitude and longitude, and when integrated over time around the orbital axis there are two
latitudes with peaks in the mass outflow rate and variations in the outflow velocity (Figs. \ref{fig:tracers}, \ref{fig:three-dimensional_density}, and \ref{fig:angle}).
The two opposite jets inflate two hot bubbles, one at each side of the equatorial plane (temperature panel in Fig. \ref{fig:4panel}).
Due to the orbital motion these bubbles stretch in the azimuthal direction and expand outwards.
The jets and the inflated bubbles lead to a concentrated equatorial outflow, despite that neither the gravity
of the secondary star nor AGB rotation were included in the simulation.

We also find that if jets are indeed launched according to the expectation from the mass accretion rate (eqs. \ref{eq:dotmaccrete} and \ref{eq:dotmjet}), they can efficiently
remove gas from the AGB envelope as the secondary star moves on a grazing orbit.
We find that even if the secondary star starts to launch jets only after it penetrated to the giant
envelope, the jets can nonetheless eject a large amount of mass from the region of the envelope outside the secondary orbit and near the equatorial plane.

The main result of the present study is that the GEE has a merit.
Our results also hint that jets might help in removing the envelope in a full common envelope evolution, i.e.,
when the secondary is deep inside the envelope rather than grazing the surface.
In a future paper we will improve our modelling by introducing a spiraling-in orbit that fully conserves the losses of energy and will follow the secondary
for longer times as it dives into the AGB.

Jets play an important role in the common envelope phase, especially during the onset of the common envelope.
Long lasting imprints of the jets, both in the GEE and in the common envelope evolution, might be a complicated morphology of the descendant nebula, whether a planetary
nebula or a nebula around a massive star.

\section*{Acknowledgments}
This work was supported by the Cy-Tera Project, which is co-funded by the European Regional Development Fund and the Republic of Cyprus through the Research Promotion Foundation.


\label{lastpage}

\begin{thebibliography}{0}
\bibitem[Armitage \& Livio(2000)]{ArmitageLivio2000} Armitage, P.~J., \& Livio, M.\ 2000, \apj, 532, 540

\bibitem[Bodenheimer \& Taam(1984)]{BodenheimerTaam1984} Bodenheimer, P., \& Taam, R.~E.\ 1984, \apj, 280, 771

\bibitem[Bondi(1952)]{Bondi1952} Bondi, H.\ 1952, \mnras, 112, 195

\bibitem[Chevalier(2012)]{Chevalier2012} Chevalier, R.~A.\ 2012, \apj, 752, L2

\bibitem[De Marco et al.(2011)]{DeMarcoetal2011} De Marco, O., Passy, J.-C., Moe, M., Herwig, F., Mac Low, M.-M., \& Paxton, B.\ 2011, \mnras, 411, 2277

\bibitem[Federrath et al.(2014)]{Federrathetal2014} {{{ {Federrath, C.,
Schr{\"o}n, M., Banerjee, R., \& Klessen, R.~S.\ 2014, \apj, 790, 128} }}}

\bibitem[Fryxell et al.(2000)]{Fryxell2000} Fryxell, B., Olson, K., Ricker, P., et al.\ 2000, \apjs, 131, 273

\bibitem[Iaconi et al.(2016)]{Iaconietal2016} Iaconi, R., Reichardt, T., Staff, J., De Marco, O., Passy, J.-C., Price, D., \& Wurster, J.\ 2016, arXiv:1603.01953

\bibitem[Ivanova \& Nandez(2016)]{IvanovaNandez2016} Ivanova, N., \& Nandez, J.~L.~A., 2016, arXiv: 1606.04923

\bibitem[Kuruwita et al.(2016)]{Kuruwitaetal2016}  Kuruwita, R.~L., Staff, J., \& De Marco, O.\ 2016, arXiv:1606.04635

\bibitem[Livio \& Soker(1988)]{LivioSoker1988} Livio, M., \& Soker, N.\ 1988, \apj, 329, 764

\bibitem[Lombardi et al.(2006)]{Lombardi2006} Lombardi, J.~C., Jr., Proulx, Z.~F., Dooley, K.~L., Theriault, E.~M., Ivanova, N., \& Rasio, F.~A.\ 2006, \apj, 640, 441

\bibitem[MacLeod \& Ramirez-Ruiz(2015)]{MacLeodRamirezRuiz2015} {{{ {MacLeod, M., \& Ramirez-Ruiz, E.\ 2015, \apj, 803, 41} }}}

\bibitem[Nandez \& Ivanova(2016)]{NandezIvanova2016} Nandez, J.~L.~A., \& Ivanova, N.\ 2016, \mnras,

\bibitem[Nandez et al.(2014)]{Nandezetal2014} Nandez, J.~L.~A., Ivanova, N., \& Lombardi, J.~C., Jr.\ 2014, \apj, 786, 39

\bibitem[Nandez et al.(2015)]{Nandezetal2015} Nandez, J.~L.~A., Ivanova, N., \& Lombardi, J.~C.\ 2015, \mnras, 450, L39

\bibitem[Nordhaus \& Spiegel(2013)]{NordhausSpiegel2013} Nordhaus, J., \& Spiegel, D.~S.\ 2013, \mnras, 432, 500

\bibitem[Ohlmann et al.(2016)]{Ohlmannetal2016} Ohlmann, S.~T., R{\"o}pke, F.~K., Pakmor, R., \& Springel, V.\ 2016, \apjl, 816, L9

\bibitem[Passy et al.(2011)]{Passy2011} Passy, J.-C., De Marco, O., Fryer, C.~L., et al.\ 2011, Evolution of Compact Binaries, Edited by Linda Schmidtobreick, Matthias R. Schreiber, and Claus Tappert. ASP Conference Proceedings, Vol. 447, 107

\bibitem[Passy et al.(2012)]{Passyetal2012} Passy, J.-C., De Marco, O., Fryer, C.~L., et al.\ 2012, \apj, 744, 52

\bibitem[Paxton et al.(2015)]{Paxtonetal2015} Paxton, B., Marchant, P., Schwab, J., et al.\ 2015, \apjs, 220, 15

\bibitem[Pudritz et al.(2012)]{Pudritzetal2012} {{{ {Pudritz, R.~E., Hardcastle, M.~J., \& Gabuzda, D.~C.\ 2012, SSRv, 169, 27 } }}}

\bibitem[Rasio \& Livio(1996)]{RasioLivio1996} Rasio, F.~A., \& Livio, M.\ 1996, \apj, 471, 366

\bibitem[Ricker \& Taam(2012)]{RickerTaam2012} Ricker, P.~M., \& Taam, R.~E.\ 2012, \apj, 746, 74

\bibitem[Sabach \& Soker(2015)]{SabachSoker2015} Sabach, E., \& Soker, N.\ 2015, \mnras, 450, 1716

\bibitem[Sandquist et al.(1998)]{SandquistTaam1998} Sandquist, E.~L., Taam, R.~E., Chen, X., Bodenheimer, P., \& Burkert, A.\ 1998, \apj, 500, 909

\bibitem[Schreier \& Soker(2016)]{SchreierSoker2016} Schreier, R., \& Soker, N.\ 2016, Research in Astronomy and Astrophysics, 16, 001

\bibitem[Shiber et al.(2015)]{Shiberetal2015} Shiber, S., Schreier, R., \& Soker, N.\ 2015, Research in Astronomy and Astrophysics, in press. arXiv:1504.04144

\bibitem[Soker(2004)]{Soker2004} Soker, N.\ 2004, \na, 9, 399

\bibitem[Soker(2013)]{Soker2013} Soker, N.\ 2013, \na, 18, 18

\bibitem[Soker(2014)]{Soker2014} Soker, N.\ 2014, arXiv:1404.5234

\bibitem[Soker(2015)]{Soker2015} Soker, N.\ 2015, \apj, 800, 114

\bibitem[Soker(2016a)]{Soker2016a} Soker, N.\ 2016, \mnras, 455, 1584

\bibitem[Soker(2016b)]{Soker2016b} Soker, N.\ 2016, \na, 47, 16

\bibitem[Staff et al.(2016a)]{Staffetal2016MN} {{{
{Staff, J.~E., De Marco, O., Macdonald, D., Galaviz, P., Passy, J.C., Iaconi, R., \& Mac Low, M.-M\ 2016a, \mnras, 455, 3511} }}}

\bibitem[Staff et al.(2016b)]{Staffetal2016} Staff, J.~E., De Marco,
 O., Wood, P., Galaviz, P., \& Passy, J.-C.\ 2016b, \mnras, 458, 832

\end{thebibliography}
\end{document}